\documentclass[conference]{IEEEtran}
\usepackage{graphicx,psfrag,epsfig,epsf,latexsym,hhline,amsmath,amssymb,multirow}
\usepackage[usenames,dvipsnames]{pstricks}
\usepackage{pst-plot}
\usepackage{pstricks-add}
\usepackage{hyperref}
\usepackage{graphicx}
\usepackage{amsthm}
\usepackage{algorithm}
\usepackage{algpseudocode}

\usepackage{blindtext}
\usepackage{etoolbox}
\IEEEoverridecommandlockouts
\graphicspath{ {figures/} }

\input{Jerry.def}

\begin{document}
\title{Lattices from Codes for Harnessing Interference: An Overview and Generalizations}
\author{Yu-Chih Huang and Krishna R. Narayanan\\
Department of Electrical and Computer Engineering \\
Texas A\&M University\\
{\tt\small {\{jerry.yc.huang@gmail.com and krn@ece.tamu.edu\}} }}

\maketitle

\begin{abstract}
In this paper, using compute-and-forward as an example, we provide an overview of constructions of lattices from codes that possess the right algebraic structures for harnessing interference. This includes Construction A, Construction D, and Construction $\pi_A$ (previously called product construction) recently proposed by the authors. We then discuss two generalizations where the first one is a general construction of lattices named Construction $\pi_D$ subsuming the above three constructions as special cases and the second one is to go beyond principal ideal domains and build lattices over algebraic integers.
\end{abstract}


\section{Introduction}
One major challenge differentiating multi-user communications from its point-to-point counterpart is that in a multi-user scenario, signals from one node would cause interference to nodes within the transmission range. Recently, there has emerged a novel perspective of dealing with interference which tries to harness interference via structured codes \cite{nazer2011CF} \cite{wilson10} \cite{ordentlich12} to name a few. The main idea behind such paradigm is to enable the destination nodes (typically relay nodes in a larger network) to compute and forward functions of messages rather than decoding them individually. The chosen functions have to in some sense match the operation induced by the channel so that the structural gains offered by the channel can be exploited.

At the heart of this strategy lies lattices constructed from codes (or nested lattice codes of Erez and Zamir \cite{erez04} to be specific). In this paper, using the compute-and-forward paradigm \cite{nazer2011CF} as an example, we aim to provide an overview of constructions of lattices from codes that are suitable for this novel interference management technique and discuss two generalizations that will expand the design space. The discussion and generalizations naturally carry over to other applications that adopt lattices from codes including integer-forcing linear receivers, compute-and-forward transform, physical-layer network-coding, etc.

In \cite{nazer2011CF}, Nazer and Gastpar adopt nested lattice codes \cite{erez04} at each node and let the relay nodes adaptively choose linear combinations of lattice codewords that are close to (a version of) the received signal. Those functions are then mapped back to linear combinations of messages in the finite field. This approach is shown to provide significant gains over the conventional strategies. Feng \textit{et al.} in \cite{Feng10} study the algebraic structure of compute-and-forward and show that the key enabler of such paradigm is the use of a ring homomorphism for mapping linear integer combinations of lattice codewords to linear combinations of messages over a finite field. They then use the isomorphism theorems in algebra to develop a general framework of constructing practical compute-and-forward schemes. In \cite{Engin12}, Tunali \textit{et al.} replace the nested lattice code in \cite{nazer2011CF} by the one constructed over Eisenstein integers and show that superior average computation rates to those in \cite{nazer2011CF} are achievable.

Both the lattice codes in \cite{nazer2011CF} and \cite{Engin12} are from Construction A \cite{LeechSloane71} \cite{conway1999sphere} which constructs lattices from linear codes. On the one hand, such construction is particularly good for compute-and-forward as lattices constructed from it possess ring homomorphisms and it is shown in \cite{erez05} to produce good lattices when the underlying linear code is over a sufficiently large prime field, properties that are required to show the results in \cite{nazer2011CF} and \cite{Engin12}. On the other hand, one major issue of lattices from Construction A is that the decoding complexity typically depends on decoding the underlying linear code. Hence, the decoding complexity can be large.

In \cite{huang13ITW} \cite{huang14}, motivated by the theory developed in \cite{Feng10}, we propose a novel multilevel lattice construction called Construction $\pi_A$ (previously called product construction) which breaks the underlying linear code into the Cartesian product of linear codes over small prime fields. Using the Chinese Remainder Theorem (CRT), we show the existence of ring homomorphisms. Moreover, we show that Construction $\pi_A$ can produce good lattices by decoding level by level so that the results in \cite{nazer2011CF} and \cite{Engin12} can be recovered by such lattices under multistage decoding, which substantially reduces the decoding complexity.

In this paper, we will first review the compute-and-forward paradigm and the algebraic structure behind it. We will then provide an overview of constructions of lattices from codes which are suitable for compute-and-forward. Particularly, we will review Construction A, Construction D, and Construction $\pi_A$. In the end, we will discuss two generalizations which enlarge the design space. Due to the space limitation, we do not provide background on lattices and abstract algebra. The reader is referred to \cite{erez05} and \cite{Hungerford74} for details.

\subsection{Notations}
Throughout the paper, we use $\mbb{N}$, $\mbb{R}$, and $\mbb{C}$ to represent the set of natural numbers, real numbers, and complex numbers, respectively. $\mbb{Z}$, $\Zi$, and $\Zw$ are the rings of integers, Gaussian integers, and Eisenstein integers, respectively. We use $\Pp(E)$ to denote the probability of the event $E$. Vectors and matrices are written in lowercase boldface and uppercase boldface, respectively. Random variables are written in Sans Serif font. We use $\times$ to denote the Cartesian product and use $\oplus$ and $\odot$ to denote the addition and multiplication operations, respectively, over a ring.

\section{Problem Statement and the Compute-and-Forward Paradigm}
\begin{figure}
    \centering
    \includegraphics[width=2.in]{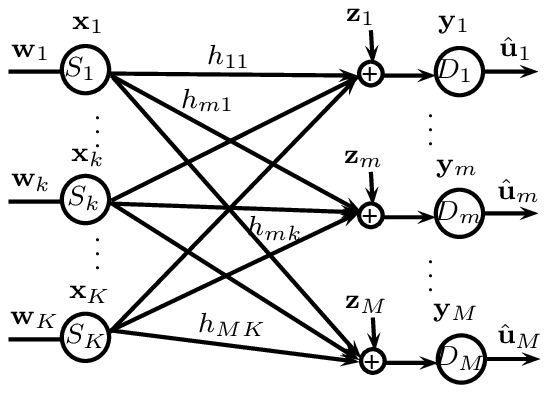}
    \caption{A compute-and-forward relay network where $S_1,\ldots,S_K$ are source nodes and $D_1,\ldots,D_M$ are destination nodes.}
    \label{fig:CF_model}
\end{figure}

In this paper, we consider the compute-and-forward network first studied by Nazer and Gastpar \cite{nazer2011CF}. There are $K$ source node and $M$ destination node and the graph is fully connected with a channel coefficient $h_{mk}\in\mbb{C}$ associated with the edge from source node $k$ to destination node $m$ as shown in Fig.~\ref{fig:CF_model}. The source node $k$ encodes its message $\mathbf{w}_k\in\mbb{F}_p^{n}$ to form a codeword $\mathbf{x}_k\in\mbb{C}^N$ which satisfies the average power constraint $\mbb{E}[\msf{X}^2]\leq P$.

The received signal at the destination node $m$ is given by
\begin{equation}\label{eqn:y_m}
    \mathbf{y}_m = \sum_{k=1}^K h_{mk}\mathbf{x}_k + \mathbf{z}_m,
\end{equation}
where $\mathbf{z}_m\sim\mc{CN}(\mathbf{0},\mathbf{I})$.

Instead of individual messages, each destination node is only interested in computing a function of messages
\begin{equation}
    \mathbf{u}_m = f_m(\mathbf{w}_1, \ldots, \mathbf{w}_K).
\end{equation}
One can think of this network model as a part of a larger network in which the destination nodes are merely relay nodes. Then computing and forwarding functions at those relay nodes mimics the behavior of random linear network coding.



In \cite{nazer2011CF}, Nazer and Gastpar proposed a novel paradigm called compute-and-forward which exploits the algebraic structure of lattices. In their framework, each source node encodes its message by an identical nested lattice code $(\Lambda_f,\Lambda_c)$ \cite{erez04}. The transmitted signal at the source node $k$ is given by
\begin{equation}
    \mathbf{x}_k = (\mathbf{t}_k -\mathbf{u}_k) \mod \Lambda_c,
\end{equation}
where $\mathbf{t}_k$ is the lattice codeword corresponding to the message $\mathbf{w}_k$ and $\mathbf{u}_k$ is a random dither.

According to the channel parameters, the destination node $m$ computes a linear combination of transmitted signals with coefficients being integers $\mathbf{a}_m=[a_{m1},\ldots,a_{mK}]$ by quantizing the following signal to the nearest lattice point in $\Lambda_f$
\begin{align}\label{eqn:NG_y}
    \mathbf{y}'_m &= \left(\alpha_m\mathbf{y}_m + \sum_{k=1}^K a_{mk}\mathbf{u}_k\right) \mod \Lambda_c \nonumber \\
    &= (\mathbf{t}_{eq,m} + \mathbf{z}_{eq,m}) \mod \Lambda_c,
\end{align}
where
\begin{equation}\label{eqn:NG_t_eq}
    \mathbf{t}_{eq,m}= \sum_{k=1}^K a_{mk}\mathbf{t}_{mk} \mod \Lambda_c,
\end{equation}
is again a lattice codeword in $(\Lambda_f,\Lambda_c)$ and
\begin{equation}\label{eqn:NG_z_eq}
    \mathbf{z}_{eq,m} = \left(\alpha_m\mathbf{z}_m + \sum_{k=1}^K (\alpha_m h_{mk}-a_{mk})\mathbf{x}_k\right),
\end{equation}
is the effective noise. A version of the decoded lattice codeword $\mathbf{t}_{eq,m}$ is then mapped to the following function via $\sigma$
\begin{equation}\label{eqn:u_NG}
    \mathbf{u}_m = b_{m1} \mathbf{w}_1\oplus\ldots\oplus b_{mK} \mathbf{w}_K,
\end{equation}
where $b_{mk}\defeq \sigma(a_{mk})\in\mbb{F}_{p}$.
It is shown in \cite{nazer2011CF} that by choosing $\alpha_m$ to be the MMSE estimator, the following computation rate is achievable
\begin{equation}\label{eqn:com_rate_m}
        R(\mathbf{h}_m,\mathbf{a}_m) = \log^+\left(\left(\|\mathbf{a}_m\|^2-\frac{P|\mathbf{h}_m^*\mathbf{a}_m|^2}{1+P\|\mathbf{h}_m\|^2}\right)^{-1}\right).
    \end{equation}
where $\log^+(.)\defeq \max\{0,\log(.)\}$.

It has been shown by Feng \textit{et al.} \cite{Feng10} that the key to relate $\mathbf{t}_{eq,m}$ a linear integer combination of codewords to $\mathbf{u}_m$ a linear combination of messages is to choose the mapping $\sigma:\mbb{Z}\rightarrow \mbb{F}_p$ a ring homomorphism and the mapping adopted in \cite{nazer2011CF} happens to be one.

\section{Some Known Constructions from Codes}
In this section, we review some known lattice constructions from codes that are suitable for compute-and-forward. i.e., constructions from codes that possess the desired homomorphisms for exploiting the structural gains offered by the channels.

\subsection{Construction A Lattices}
\begin{figure}
    \centering
    \includegraphics[width=2.5in]{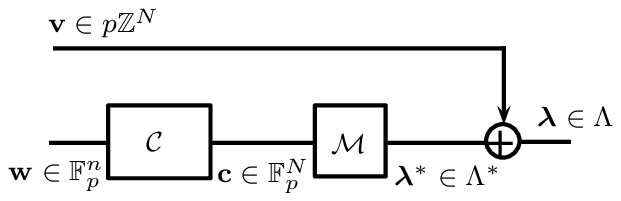}
    \caption{Construction A.}
    \label{fig:const_A_Z}
\end{figure}

We review the Construction A lattices over $\mbb{Z}$ and discuss some properties of such lattices and some related constructions. A depiction of Construction A can be found in Fig.~\ref{fig:const_A_Z}.

{\bf \underline{Construction A}} \cite{LeechSloane71} \cite{conway1999sphere} Let $p$ be a prime. Let $n$, $N$ be integers such that $n\leq N$ and let $\mathbf{G}$ be a generator matrix of an $(N,n)$ linear code over $\mbb{F}_p$. Construction A consists of the following steps:
\begin{enumerate}
    \item Define the discrete codebook $C=\{\mathbf{x}=\mathbf{G}\odot\mathbf{w}:\mathbf{w}\in\mbb{F}_p^n\}$ where all operations are over $\mbb{F}_p$.
    \item Construct $\Lambda^*\defeq \mc{M}(C)$ where $\mc{M}:\mbb{F}_p\rightarrow \mbb{Z}/p\mbb{Z}$ is a ring isomorphism. (For $\mbb{Z}$, the natural mapping suffices.)
    \item Tile $\Lambda^*$ to the entire $\mbb{R}^N$ to form $\Lambda_{\text{A}}\defeq \Lambda^* + p\mbb{Z}$.
\end{enumerate}
Note that the existence of the ring isomorphism in step 2) is guaranteed since $p\mbb{Z}$ is a maximal ideal. It can be shown that a real vector $\boldsymbol\lambda$ belongs to $\Lambda_{\text{A}}$ if and only if $\sigma(\boldsymbol\lambda)\in C$ where $\sigma\defeq \mc{M}^{-1}\circ\hspace{-3pt}\mod p\mbb{Z}$ is a ring homomorphism.

After a long pursuit by pioneers like de Buda, Poltyrev, Loeliger, Forney, Rimoldi, Urbanke, and etc., Erez and Zamir in 2004 finally showed that lattice codes constructed from Construction A can achieve the AWGN capacity under lattice decoding (see \cite{erez04} and the reference therein) as $p\rightarrow\infty$. Such lattice codes are particularly suitable for compute-and-forward as after decoding the fine lattice point, one can use the ring homomorphism $\sigma$ to obtain the corresponding linear combination of messages over $\mbb{F}_p$.

Extensions of Construction A to other PIDs such as Gaussian integers $\Zi$ and Eisenstein integers $\Zw$ are possible \cite{conway1999sphere}. In \cite{Engin12}, following \cite{erez05}, Tunali $\textit{et al.}$ showed the goodness of Construction A lattices over $\Zw$. Motivating by the fact that $\Zw$ quantizes the complex field better than $\Zi$ does, they then used lattice codes generated from Construction A over $\Zw$ for compute-and-forward. Constructing practical ensembles of lattices from Construction A is also an active research area. Building upon non-binary low-density parity check (LDPC) codes, di Pietro \textit{et al.} \cite{diPietro12} (and Tunali $\textit{et al.}$ \cite{Engin12} as well) propose the low-density A (LDA) lattice ensemble from Construction A and show that such lattices can achieve the Poltyrev-limit under maximal likelihood decoding \cite{diPietro13}. Tunali $\textit{et al.}$ in \cite{Engin13ITW} further replace LDPC codes by spatially-coupled LDPC codes and present a BP-threshold of 0.19 dB from the Poltyrev-limit at a block length of $1.29\times 10^6$.


\subsection{Construction D Lattices}
We now consider the Construction D lattices \cite{BarnesSloane83} \cite[Page 232]{conway1999sphere}. Let $C^1 \subseteq C^2 \subseteq  \ldots \subseteq C^{L+1}$ be a sequence of nested linear codes over $\mbb{F}_p$ where $C^{L+1}$ is the trivial $(N,N)$-code and $C^l$ is a $(N,n^l)$-code for $l\in\{1,2,\ldots r\}$ with $n^1 \leq \ldots \leq n^r$. The codes are guaranteed to be nested by choosing $\{ \mathbf{g}_1,\ldots,\mathbf{g}_N \}$ which spans $C^{L+1}$ and then using the first $n^l$ vectors $\{ \mathbf{g}_1,\ldots,\mathbf{g}_{n^l} \}$ to generate $C^l$. We are now ready to state Construction D of lattices.

{\bf \underline{Construction D}} A lattice $\Lambda_{\text{D}}$ generated by Construction D with $L+1$ level is given as follows.
\begin{equation}
    \Lambda_{\text{D}} = \bigcup \left\{ p^L \mbb{Z}^N + \sum_{1\leq l \leq L} p^{l-1} \sum_{1\leq i \leq n^l} a_{li} \mathbf{g}_i |a_{li}\in\mbb{F}_p \right\},
\end{equation}
where all the operations are over $\mbb{R}^N$. Similar to Construction A, extensions of Construction D to other PIDs such as $\Zi$ and $\Zw$ are possible.

It was shown by Forney \cite{forney2000} that Construction D lattices with any fixed $p$ and sufficiently large $L$ can achieve the Poltyrev-limit under multistage decoding. Thus, one can choose $p=2$ and always work with the binary field. Therefore, in general, the decoding complexity of Construction D lattices is much smaller than that of Construction A lattices. There have been several attempts to construct practical ensemble of lattices from Construction D. In \cite{sakzad10}, Sakzad \textit{et al.} proposed the turbo lattices from Construction D together with turbo codes. Although no theoretical proofs showing the ability of achieving the Poltyrev-limit, simulation results reported that turbo lattices can approach the Poltyrev-limit to within 0.5 dB at $p_e=10^{-5}$ with the code length roughly 10000. In \cite{YanLingWu13}, Yan \textit{et al.} constructed sequences of nested polar codes and used them in conjunction with Construction D to generate polar lattices. Similar to the linear code counterpart (i.e., the polar codes), such lattices can be shown achieving the Poltyrev-limit and explicit constructions of good polar lattices are possible. Very recently, Vem \textit{et al.} in \cite{vem14} proposed a means to construct sequences of nested spatially-coupled LDPC codes and adopted Construction D to construct the so-called spatially-coupled LDPC lattices. This ensemble of lattices is shown to achieve the Poltyrev-limit under \textit{belief propagation decoding}.

To use Construction D lattices for compute-and-forward, there are some challenges that need to be conquered. First of all, to the best of our knowledge, there is no proof showing that Construction D can produce lattices that are good for MSE quantization. Therefore, efficient shaping techniques are called for. Secondly, unlike Construction A lattices, mapping linear integer combinations of lattice points to linear combinations of codewords over finite field is not an easy task. Hence, lattices from Construction D may not be straightforwardly applied to compute-and-forward if one insists on coding over finite field. On the other hand, if we are allowed to work over a finite-chain-ring, the second issue can be circumvented as Construction D lattices can be deemed as Construction A lattices with linear codes over finite-chain-rings \cite[Proposition 2]{feng11rings} and one can relate integer linear combinations of lattice points to linear combinations of codewords over finite-chain-rings. But then one has to pay extra attention to zero divisors. 

\subsection{Construction $\pi_A$ Lattices}\label{sec:const_pi_A}
\begin{figure}
    \centering
    \includegraphics[width=2.5in]{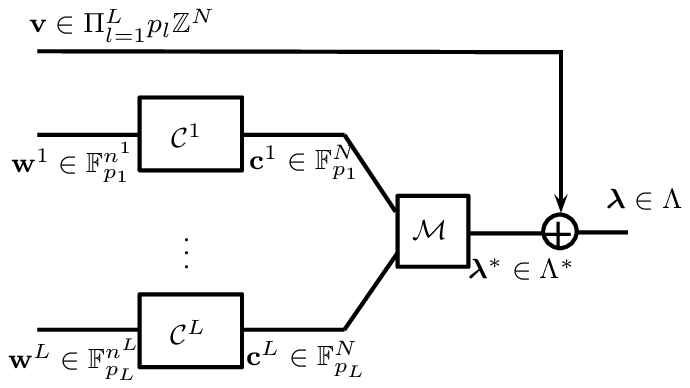}
    \caption{Construction $\pi_A$.}
    \label{fig:lattice_const}
\end{figure}

Motivated by the problems occurring when using Construction A and D lattices for compute-and-forward, the authors proposed a novel lattice construction called Construction $\pi_A$ \cite{huang13ITW} \cite{huang14} shown in Fig.~\ref{fig:lattice_const}. This construction is built upon the algebraic foundation established by Feng \textit{et al.} \cite{Feng10}.

\textbf{\underline{Construction $\pi_A$}} Let $p_1, p_2,\ldots, p_L$ be primes which are relatively prime. Let $n^l$, $N$ be integers such that $n^l\leq N$ and let $\mathbf{G}^l$ be a generator matrix of an $(N,n^l)$ linear code over $\mbb{F}_{p_l}$ for $l\in\{1,\ldots,L\}$. Construction $\pi_A$ consists of the following steps,

\begin{enumerate}
    \item Define the discrete codebooks $C^l=\{\mathbf{x}=\mathbf{G}^l\odot\mathbf{w}^l:\mathbf{w}^l\in\mbb{F}_{p_l}^{n^l}\}$ for $l\in\{1,\ldots,L\}$.
    \item Construct $\Lambda^* \defeq \mc{M}(C^1,\ldots,C^L)$ where $\mc{M}:\times_{l=1}^L \mbb{F}_{p_l}\rightarrow \mbb{Z}/\Pi_{l=1}^L p_l \mbb{Z}$ is a ring isomorphism.
    \item Tile $\Lambda^*$ to the entire $\mbb{R}^N$ to form $\Lambda_{\pi_A} \triangleq  \Lambda^* + \Pi_{l=1}^L p_l \mbb{Z}^N$.
\end{enumerate}
Note that the existence of the ring isomorphism in step 2) is guaranteed by CRT. It can be shown that a real vector $\boldsymbol\lambda$ belongs to $\Lambda_{\pi_A}$ if and only if $\sigma(\boldsymbol\lambda)\in C^1\times\ldots\times C^L$ where $\sigma\defeq \mc{M}^{-1}\circ\hspace{-3pt}\mod \Pi_{l=1}^L p_l\mbb{Z}$ is a ring homomorphism.

It has been shown that Construction $\pi_A$ is able to generate lattices that are simultaneously good for MSE quantization and Poltyrev good under multistage decoding \cite{huang14}. This means that the complexity of such lattices is only dominated by decoding the linear code whose field size is equal to the largest prime $\max_l p_l$ in the prime factorization rather than $\Pi_{l=1}^L p_l$. Unlike Construction A letting $p$ tend to infinity and Construction D letting $L$ tend to infinity, Construction $\pi_A$ let $\Pi_{l=1}^L p_l$ tend to infinity and allows one to play with these parameters. Moreover, in Construction $\pi_A$, the linear code in each level does not have to be nested in other linear codes; hence, the rate allocation is much easier than Construction D.

We further apply the Construction $\pi_A$ lattices to compute-and-forward and proposed the \textit{multistage} compute-and-forward scheme. The main enabler is to realize that any integer $a\in\mbb{Z}$ can be decomposed as $a = \mc{M}(b^1,\ldots,b^L)+\Pi_{l=1}^L p_l \tilde{a},$
where $\tilde{a}\in\mbb{Z}$ and $b^l\in\mbb{F}_{p_l}$ that can be obtained from $\sigma(a)$. i.e., any integer can be represented as its coordinate in $\times_{l=1}^L \mbb{F}_{p_l}$. The transmitter $k$ first splits its message into $L$ streams $\mathbf{w}_k^1,\ldots,\mathbf{w}_k^L$ where $\mathbf{w}_k^l\in\mbb{F}_{p_l}^{n^l}$. Each stream is then separately encoded by the linear code in that level. The encoder gathers all the coded streams and use $\mc{M}$ to map them to the constellation.

At the relay $m$, instead of \eqref{eqn:u_NG}, we opt to decode
\begin{equation}\label{eqn:u_MLCF}
    \mathbf{u}_m^l \defeq b_{m1}^l\odot \mathbf{w}_1^l \oplus\ldots\oplus b_{mK}^l\odot \mathbf{w}_K^l,
\end{equation}
for all $l\in\{1,\ldots,L\}$ level by level. In \cite{huang14}, we showed that the multistage compute-and-forward achieves the same computation rates as those in \cite{nazer2011CF} (see \eqref{eqn:com_rate_m} as well) with multistage decoding, which substantially decreases the decoding complexity.

\section{Generalizations}\label{sec:general}
In this section, we generalize the above lattice constructions in the following two aspects.
The first generalization discussed in this section is a novel lattice construction called Construction $\pi_D$. This construction is a direct consequence of CRT and subsumes Construction A, Construction D, and Construction $\pi_A$ as special cases. This will substantially expand the design space. Another thing one may have noticed is that all the lattices constructed so far are over PIDs such as $\mbb{Z}$, $\Zi$, and $\Zw$. Mathematicians have known how to construct lattices over other commutative rings for a long time (see for example \cite{boutros96} \cite{Kosit} and reference therein). The second generalization is to go beyond PIDs and allow one to build lattices over rings of algebraic integers. This will result in increased computation rates for some channel realizations.

\subsection{Construction $\pi_D$}\label{sec:pi_D}
Let $q\in\mbb{N}$ be any natural number whose prime factorization is given by $q=\Pi_{l=1}^L p_l^{e_l}$. From CRT, there exists a ring isomorphism $\mc{M}:\times_{l=1}^L \mbb{Z}_{p_l^{e_l}}\rightarrow \mbb{Z}/ q \mbb{Z}$. Moreover, $\sigma\defeq \mc{M}^{-1}\circ \hspace{-3pt}\mod q\mbb{Z}$ is a ring homomorphism.

\textbf{\underline{Construction $\pi_D$}} Let $q\in\mbb{N}$ whose prime factorization is given by $q=\Pi_{l=1}^L p_l^{e_l}$. Let $n^l$, $N$ be integers such that $n^l\leq N$ and let $\mathbf{G}^l$ be a generator matrix of an $(N,n^l)$ linear code over $\mbb{Z}_{p_l^{e_l}}$ for $l\in\{1,\ldots,L\}$. Construction $\pi_D$ consists of the following steps,

\begin{enumerate}
    \item Define the discrete codebooks $C^l=\{\mathbf{x}=\mathbf{G}^l\odot\mathbf{w}^l:\mathbf{w}^l\in(\mbb{Z}_{p_l^{e_l}})^{n^l}\}$ for $l\in\{1,\ldots,L\}$.
    \item Construct $\Lambda^* \defeq \mc{M}(C^1,\ldots,C^L)$ where $\mc{M}:\times_{l=1}^L \mbb{Z}_{p_l^{e_l}}\rightarrow \mbb{Z}/ q \mbb{Z}$ is a ring isomorphism.
    \item Tile $\Lambda^*$ to the entire $\mbb{R}^N$ to form $\Lambda_{\pi_D} \triangleq  \Lambda^* + q \mbb{Z}^N$.
\end{enumerate}
Similar to $\Lambda_{\pi_A}$, it can be shown that a real vector $\boldsymbol\lambda$ belongs to $\Lambda_{\pi_D}$ if and only if $\sigma(\boldsymbol\lambda)\in C^1\times\ldots\times C^L$ where $\sigma\defeq \mc{M}^{-1}\circ\hspace{-3pt}\mod q\mbb{Z}$ is a ring homomorphism.

Note that when setting $L=1$ and $e_1=1$, Construction $\pi_D$ reduces to Construction A over a finite field $\mbb{F}_{p_1}$. Setting $L=1$ makes it Construction A over a finite-chain-ring $\mbb{Z}_{p_1^{e_1}}$, which subsumes Construction D as a special case. Finally, when setting $e_1=\ldots=e_L=1$, we obtain Construction $\pi_A$. Hence, Construction $\pi_D$ is a general means of constructing lattices from codes and contains Construction A, Construction D, and Construction $\pi_A$ as special cases. Moreover, in Construction $\pi_D$, $q$ can take any natural number regardless its prime factorization. Thus, the proposed construction substantially expands the design space and further eases the rate allocation problem.


To show the ability to produce Poltyrev good lattices, for the proposed construction, one can follow the proof in \cite{huang14} with a careful treatment to those levels with $e_l\neq 1$. One option is to use Construction D for those levels, i.e., one uses a sequence of $e_l$ nested linear codes to construct a linear code over $\mbb{Z}_{p_l^{e_l}}$. Another option is to adopt a capacity-achieving linear code over $\mbb{Z}_{p_l^{e_l}}$ proposed in \cite{HitronErez12} at the $l$th level. On the other hand, for the goodness for MSE quantization, the proposed construction may suffer the same fate as Construction D unless we enforce $e_1=\ldots=e_L=1$ (Construction $\pi_A$).


\subsection{Lattices over Algebraic Integers}\label{sec:algebraic}
We particularly look at constructing lattices over rings of algebraic integers with degree 2, namely the quadratic integers. The reasons that we pick such rings are twofold. First of all, the channel coefficients we are trying to quantize lie in $\mbb{C}$, which is an extension field of $\mbb{R}$ with degree 2. Hence, it makes perfect sense to first investigate extensions with degree 2. Secondly, quadratic fields have been extensively studied and many properties have been discovered. This makes the generalization a lot easier. Please consult \cite{lang94} for background knowledge on algebraic number theory.

In a nutshell, a number field $\mbb{K}$ is a finite extension of $\mbb{Q}$ and its ring of integers is $\mfk{O}_{\mbb{K}}=\mbb{K}\cap\mbb{A}$ where $\mbb{A}$ is the ring of all algebraic integers. Any quadratic field can be expressed as $\mbb{Q}(\sqrt{d})$ with $0,1\neq d\in\mbb{Z}$ square-free. Its ring of integers is $\Ok=\mbb{Z}[\xi]$ where
\begin{equation}\label{eqn:zbasis}
        \xi = \left\{\begin{array}{ll}
        \sqrt{d},                                           & d\equiv 2,3\mod 4, \\
        \frac{1+\sqrt{d}}{2},                                    & d\equiv 1\mod 4.\\
        \end{array} \right.
\end{equation}
Note that when $d=-1$ and $d=-3$, we have $\Zi$ and $\Zw$, respectively. Not every $\mbb{Z}[\xi]$ forms a PID. In fact, for imaginary quadratic fields, there are exactly 9 of these are PIDs, which correspond to $d\in\{-1,-2,-3,-7,-11,-19,-43,-67,-163\}$. Fortunately, for such rings, there is a systematic way to identify prime ideals. Moreover, every prime ideal $\mfk{p}\in\Ok$ lies above $p$ is maximal in $\Ok$; hence, we still have the property that $\Ok/\mfk{p}\cong \mbb{F}_{p^f}$ where $f$ is the inertial degree.

\begin{example}\vspace{-5pt}
    Consider $\mbb{Q}[\sqrt{-15}]$ whose ring of integers is $\Ok=\mbb{Z}[\xi]$ with $\xi = \frac{1+\sqrt{-15}}{2}$. This is not a PID. One can show that $17\Ok$ splits into two prime ideals, namely $17\Ok = \mfk{p}\bar{\mfk{p}}$ where $\mfk{p}=(17,6+\sqrt{-15})$. Moreover, we have $\Ok/\mfk{p}\cong\mbb{F}_{17}$. In Fig.~\ref{fig:Z_15_homo}, we show the coset decomposition and the corresponding ring isomorphism.

    \begin{figure}
    \centering
    \includegraphics[width=2.5in]{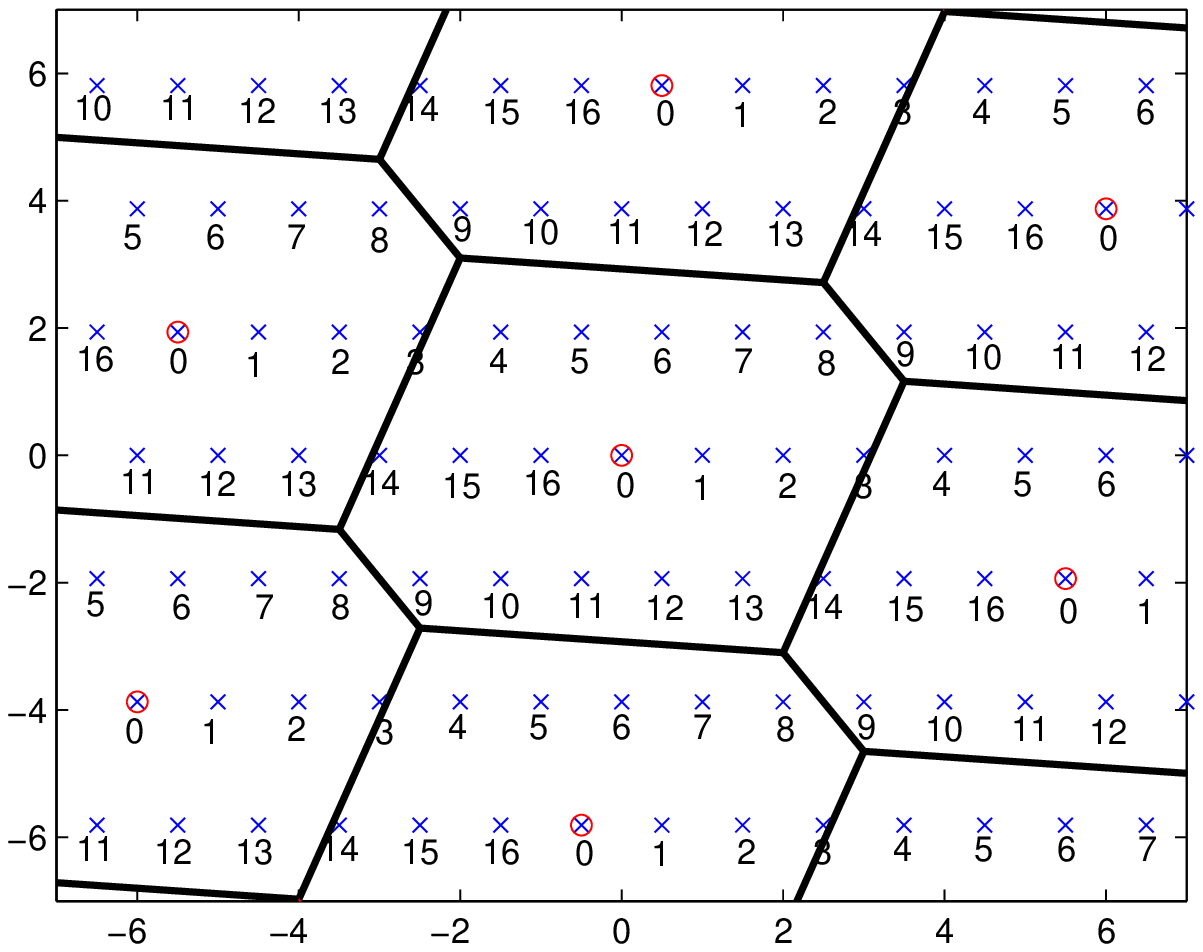}
    \vspace{-15pt}
    \caption{Coset decomposition of $\Ok/\mfk{p}$ and the corresponding ring isomorphism $\mc{M}:\mbb{F}_{17}\rightarrow\Ok/\mfk{p}$.}
    \label{fig:Z_15_homo}
    \vspace{-15pt}
\end{figure}
\end{example}
\vspace{-5pt}
In what follows, we only present Construction A over $\Ok$ but its extensions to other constructions are possible.

{\bf \underline{Construction A over $\Ok$}} Let $n$, $N$ be integers such that $n\leq N$ and let $\mathbf{G}$ be a generator matrix of an $(N,n)$ linear code over $\mbb{F}_{p^f}$. Construction A over $\Ok$ consists of the following steps:
\begin{enumerate}
    \item Define the discrete codebook $C=\{\mathbf{x}=\mathbf{G}\odot\mathbf{y}:\mathbf{y}\in\mbb{F}_{p^f}^n\}$ where all operations are over $\mbb{F}_{p^f}$.
    \item Construct $\Lambda^*\defeq \mc{M}(C)$ where $\mc{M}:\mbb{F}_{p^f}\rightarrow \Ok/\mfk{p}$ is a ring isomorphism.
    \item Tile $\Lambda^*$ to the entire $\mbb{C}^N$ to form $\Lambda\defeq \Lambda^* + \mfk{p}^N$.
\end{enumerate}

We expect to show the goodness of such lattices and to benefit from it in the scenario where we have feedback from destination nodes by letting source nodes to choose the best ring of integers to work with. This framework can also be incorporated with the phase precoded approach in \cite{sakzad14} to further improve the performance.

\vspace{-5pt}
\bibliographystyle{ieeetr}
\bibliography{journal_abbr,prod}

\end{document}